\documentclass{article}
\usepackage{ijcai24}

\usepackage{times}
\usepackage{soul}
\usepackage{amssymb}
\usepackage[breakable]{tcolorbox}
\usepackage{url}
\usepackage[hidelinks]{hyperref}
\usepackage[utf8]{inputenc}
\usepackage[small]{caption}
\usepackage{graphicx}
\usepackage{amsmath}
\usepackage{csquotes}
\usepackage{tabularx}
\usepackage{amsthm}
\usepackage{booktabs}
\usepackage{algorithm}
\usepackage{algorithmic}
\usepackage[switch]{lineno}

\title{From Skepticism to Acceptance: \\Simulating the Attitude Dynamics Toward Fake News}

\author{
Yuhan Liu$^1$
\and
Xiuying Chen$^{2*}$ \and
Xiaoqing Zhang$^{1}$\and
Xing Gao$^3$\and
Ji Zhang$^{3}$\and
Rui Yan$^{1}$\thanks{\quad Corresponding authors.}\\
\affiliations
$^1$Gaoling School of Artificial Intelligence, Renmin University of China\\
$^2$Mohamed bin Zayed University of Artificial Intelligence\\
$^3$Alibaba DAMO Academy\\
\emails
yuhan.liu@ruc.edu.cn
}

\begin{document}

\maketitle

\begin{abstract}
In the digital era, the rapid propagation of fake news and rumors via social networks brings notable societal challenges and impacts public opinion regulation.  
Traditional fake news modeling typically forecasts the general popularity trends of different groups or numerically represents opinions shift.
However, these methods often oversimplify real-world complexities and overlook the rich semantic information of news text. 
The advent of large language models (LLMs) provides the possibility of modeling subtle dynamics of opinion.
Consequently, in this work, we introduce a Fake news Propagation Simulation framework (FPS) based on LLM, which studies the trends and control of fake news propagation in detail. 
Specifically, each agent in the simulation represents an individual with a distinct personality. 
They are equipped with both short-term and long-term memory, as well as a reflective mechanism to mimic human-like thinking. 
Every day, they engage in random opinion exchanges, reflect on their thinking, and update their opinions.
Our simulation results uncover patterns in fake news propagation related to topic relevance, and individual traits, aligning with real-world observations.
Additionally, we evaluate various intervention strategies and demonstrate that early and appropriately frequent interventions strike a balance between governance cost and effectiveness, offering valuable insights for practical applications.
Our study underscores the significant utility and potential of LLMs in combating fake news\footnote{We have released the code and appendix at \url{https://github.com/LiuYuHan31/FPS}}.
\end{abstract}

\section{Introduction}
Online social media provides an accessible and cost-effective way to share information, promoting quick knowledge dissemination. 
However, its widespread use also leads to misinformation, causing global panic and emphasizing the importance of controlling false information.
For example, during the 2016 US presidential election, fake news comprised approximately 6\% of total news consumption~\cite{grinberg2019fake}. 
This issue is not confined to politics; it extends to other sectors like the stock markets~\cite{bollen2011twitter}, the aftermath of terrorist attacks~\cite{starbird2014rumors}, and responses to natural disasters~\cite{gupta2013faking}.

\begin{figure*}[tb]
    \centering
    \includegraphics[width=1\linewidth]{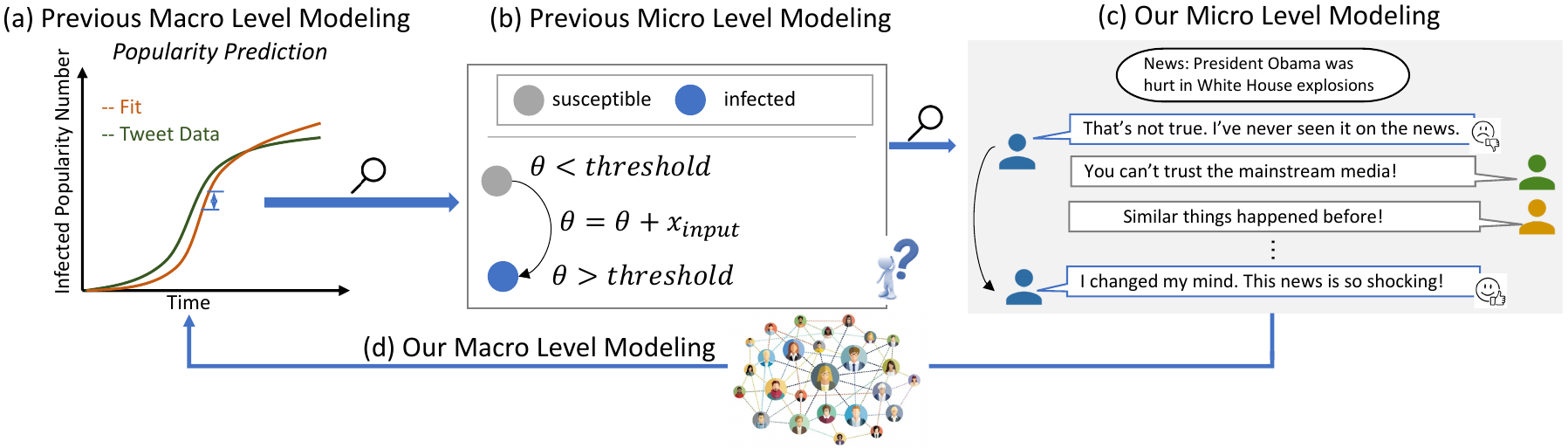}
    \vspace{-6mm}
    \caption{(a) Previous macro-level fake news modeling focused on predicting the overall infected population but lacked a detailed analysis of the dynamics in human attitudes. 
(b) Previous micro-level models translated human opinions and communication into numerical values, i.e., $\theta \in \mathbb{R}$ and $x \in \mathbb{R}$. 
(c) Our micro-level simulation uniquely captures attitude changes through natural language processing. 
(d) Additionally, our multiple agents constitute a macro-level simulation that also enables the prediction of popularity trends.
    }
    \label{fig:intro}
\end{figure*}

A variety of modeling methods have been developed to study the mechanisms behind the propagation of fake information~\cite{garimella2017balancing,wang2019efficient}.
From the macro-level, \cite{kimura2009efficient} categorize populations into susceptible and infected groups, and define transformation probabilities for each group to simulate the macro-level propagation mechanism, as depicted in Figure~\ref{fig:intro}(a).

More detailed, from the micro-level, \cite{jalili2017information} define the numerical conditions that determine whether each individual will change their opinion, as illustrated in Figure~\ref{fig:intro}(b). 
However, these models commonly depend on numerical representations for opinions and messages. 
Such a simplified approach often fails to capture the complex linguistic nuances found in real-life conversations. 
For instance, the intricate reasoning process, thoughts, and opinions about a topic ought not to be merely reduced to a single sentiment score. 
Additionally, individuals with diverse personality traits have varied reactions to the same subject, which cannot be accurately represented by these numerical models.

In this work, we propose a Fake news Propagation Simulation framework (FPS), with each individual in the network represented as an LLM agent.
This method offers several advantages. 
First, FPS allows for the simulation of users with varied personas and backgrounds, enabling researchers to study diverse behavioral patterns. 
Second, LLM-based simulations effectively replicate the textual nature of fake news, complex human reasoning, and dynamic opinion shifts, thereby enhancing explainability.
Third, the scalability of LLM-based simulations enables the analysis of fake news propagation across diverse scenarios and demographic groups, thus offering extensive and valuable insights.

Specifically, in FPS, we initialize agents in the network with unique personas including age, name, educational background, and personal traits.
The propagation of information begins with an infected individual who believes in the fake news and communicates this belief to others. 
On each day, each agent randomly communicates with several other agents and changes their opinions on the topic, deciding whether to believe in the fake news based on their reasoning and prior interactions. 
Each agent is equipped with a short-term memory to capture the day's interactions and a long-term memory for broader context, along with a reflective reasoning process to mimic the human thought process.

Furthermore, we introduce an official agent with different intervention mechanisms to counter the propagation of fake news. 
From a macro-level perspective, we calculate the overall popularity of the infected, susceptible, and recovered populations to understand broader trends. 
Concurrently, our micro-level analysis concentrates on tracking the evolving viewpoints of each individual.

Our FPS is validated through comprehensive simulation experiments, closely aligning with real-world observations from prior research. 
Notably, our findings show that political fake news prpagates notably faster than topics such as terrorism, natural disasters, science, urban legends, or financial information, consistent with previous studies~\cite{vosoughi2018spread}.
Moreover, agents characterized by specific traits are more susceptible to believing in fake news~\cite{ibrahim2022effects,mirzabeigi2023role,afassinou2014analysis}. 

From a governance perspective, our findings show that addressing fake news just once is not enough. 
Early and consistent efforts to correct misinformation work best in maintaining low propagation of fake news, offering important guidance for timely and effective information management.

Our contribution can be summarized in the following ways:
Firstly, we developed an FPS framework based on LLM for fake news, offering extensive semantic information and analysis material for macro- and micro-level studies on fake news.
Second, our experiments align closely with conclusions drawn from real-world studies, confirming the value of our FPS as a research tool.
Finally, we demonstrate the effectiveness of early and frequent interventions in mitigating the propagation of fake news, providing suggestions for policy formulation and public awareness.

\section{Related Work}

\textbf{Fake News Detection.}
Fake news detection is an essential step in the fight against misinformation~\cite{guo2021does}. 
Earlier works in this area include the study by \cite{qian2018neural}, which focuses on the early detection of fake news, considering only the text of news articles available at the time of detection.  
As the field evolved, a variety of methods have been introduced to improve detection efficacy. 
For example, \cite{jin2022towards} move towards fine-grained reasoning for fake news detection by better reflecting the logical processes of human thinking and enabling the modeling of subtle clues. 
Our research extends these foundations by tackling the propagation of fake news post-detection.

\begin{figure*}[tb]
    \centering
    \includegraphics[width=0.9\linewidth]{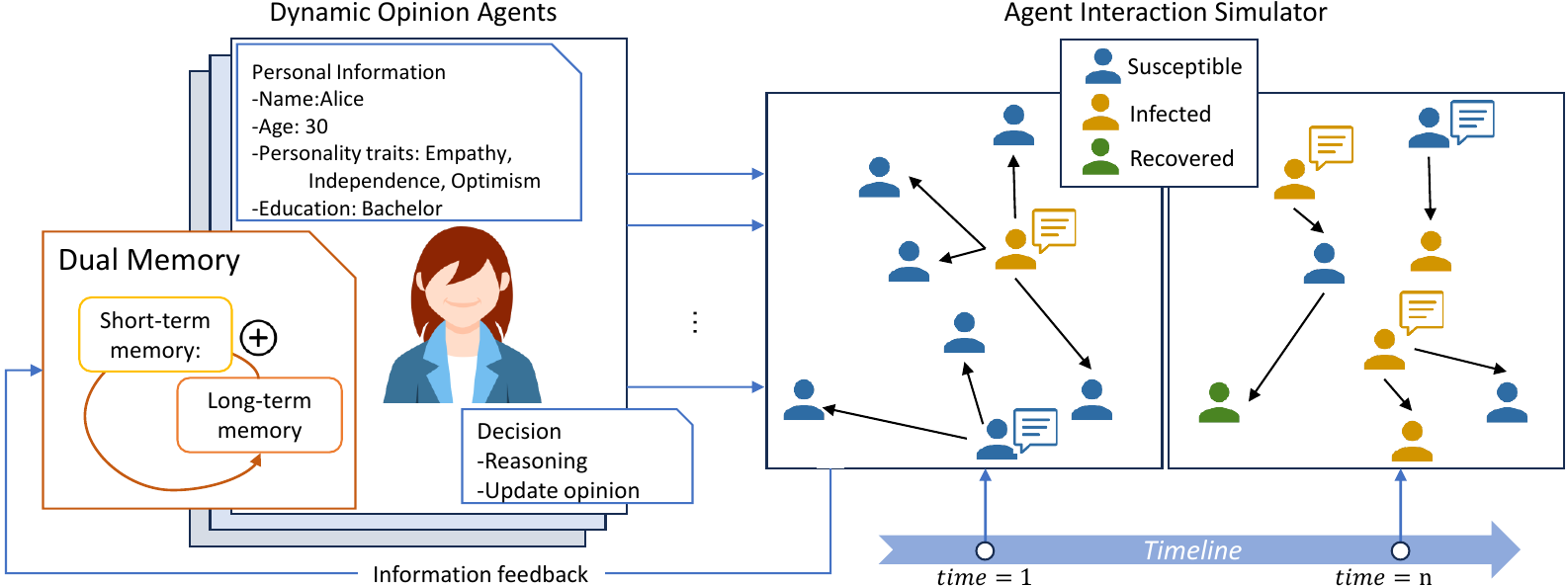}
    \caption{Our framework equips each agent with reasoning and response capabilities by creating a feedback loop between dynamic opinion agents (DOA) and an agent interaction simulator (AIS).
   `Susceptible' denotes individuals skeptical of the fake news, `infected' refers to those who believe in the fake news, and `recovered' means individuals who previously believed the fake news but now do not.}
    \label{fig:model}
\end{figure*}

\textbf{Fake News Propagation Modeling.}
Fake information propagation modeling plays a key role in understanding misinformation propagation and is vital for interventions like early warning~\cite{garimella2017balancing}, information blocking~\cite{song2015node}, and truth confrontation~\cite{wang2019efficient}.
These models generally fall into three categories~\cite{sun2023fighting}: Epidemic models such as the SIR (Susceptible-Infected-Recovered) model~\cite{zhu2017rumor} divide the population into different groups and define the transformation probabilities of each group to simulate the propagation mechanism at the macro level.
Point process-based methods treat information propagation as a stochastic arrival process, thus modeling the arrival of propagation~\cite{chen2019information,gao2019taxonomy}.
Diffusion models define the propagation conditions of each individual based on a numerical threshold~\cite{jalili2017information}. 
Unlike previous studies, our research adopts an LLM-based simulation method, offering a textual approach instead of traditional numerical calculations.

\textbf{LLM-based Agents for Social Simulation.}
Integrating LLMs into simulating social dynamics represents a burgeoning field of research, yielding promising results~\cite{park2023generative,kaiya2023lyfe,li2023quantifying}. 
These LLM-based generative agents excel in digital environments, demonstrating proficiency in natural language tasks~\cite{chen2023improving,chen2023topic}.
~\cite{tornberg2023simulating} used LLMs and agent-based modeling to simulate social media, focusing on news feed algorithms and offering real-world insights.
Further, \cite{park2022social} demonstrated that LLM-based agents can generate social media content indistinguishable from that produced by humans. 
Our approach of using LLM for agent-based fake news simulation is, to our knowledge, a pioneering effort in this field.

\section{Method}

\subsection{Problem Formulation}
Formally, we construct a simulation with a pool of $N$ LLM agents $\mathcal{A} =( a_1, \ldots, a_N)$ and a fake news topic $F$. 
For initialization, each agent has a unique persona, including their initial attitude towards the fake news.
On $t$-th day, each agent $a_i$ will randomly interact with $c$ other agents from the pool $\mathcal{A}$. 
At the end of the day, every agent reflects on the exchanged information and decides whether to believe in the fake news. 
This process is iterated over $T$ days. 
Through these daily iterations, agent opinions are accumulated to plot the trajectory of different populations, resulting in a curve that illustrates the dynamics of the fake news within the network.
Additionally, we scrutinize the evolution of the agents' beliefs to examine how individual and collective opinions shift over time.

\subsection{Our Simulation Framework}
As depicted in Figure~\ref{fig:model}, our framework FPS integrates two components: a Dynamic Opinion Agent (DOA) to simulate each agent's cognitive processes and an Agent Interaction Simulator (AIS) to construct the interaction environment.
Within the DOA module, each agent's decision-making is powered by LLM, with a predefined role that includes attributes such as education level, gender, and personality traits.
Daily, agents engage in discussions with their peers, reflecting on these interactions and adjusting their beliefs on the fake news accordingly. 
The AIS module plans the encounters, determining which agents interact, and the frequency of these interactions per day. 
Additionally, in scenarios where official announcements are released to clarify the fake news, the AIS takes charge of distributing this information. 
At the end of each day, the simulation progresses by one step, and the agents' belief states are updated.

\subsection{Dynamic Opinion Agent}
The DOA agent focuses on the micro-level, where the dynamics of each agent's opinions can be studied in detail. 

\textbf{Persona.}
We randomly equip each agent with persona $p_i$ including their name, age, trait, and education level, since these are factors that might influence their attitude toward fake news.
When designing the traits, we adhere to the Big 5 trait model~\cite{barrick1991big}. 
This model is widely recognized for its effectiveness in encapsulating key personality dimensions.

\textbf{Dual Memory.}
In our model, we consider that an individual's opinion is influenced not only by their own beliefs but also by their interactions with others. 
This interaction-driven change in thought is gradual and cumulative, rather than immediate. 
Accordingly, in our simulation, agents engage with a random number of others' opinions each day, leading to a periodic update of their views.
\begin{algorithm}[tb]
\caption{Fake News Propagation Simulation}
\label{alg:generative_agent}
\begin{algorithmic}[1]
\STATE \textbf{Input:} Number of agents $N$, interaction rate $c$
\STATE \textbf{Output:} Final memory state and opinion of each agent
\STATE \textbf{Initialize agents:}
\FOR{each agent $i$ in $1$ to $N$}
    \STATE Randomly assign persona $p_i$
    \STATE Define long-term memory $m^l_{i,0}$ and short-term memory $m^s_{i,0}$
    \STATE Set initial belief state $o^b_{i,0}$ and initial text opinion descriptor $o^l_{i,0}$ 
\ENDFOR
\STATE \textbf{Simulate daily interactions:}
\FOR{each day $t$ in $1$ to $T$}
    \FOR{each agent $i$}
        \STATE Select $c$ agents to interact with $(a_1,...,a_{c})$
        \STATE Write $i$-th agent's short-term memory $m^s_{i,t}$ with details from the day's interactions
        \STATE Based on $m^s_{i,t}$ and long-term memory $m^l_{i,t}$ update long-term memory: $m^l_{i,t}=f^l_m(m^s_{i,t},m^l_{i,t-1})$ 
        \STATE Based on $o^l_{i,t-1},m^l_{i,t},p_i$, update $o^b_{i,t},o^l_{i,t}=f_o(p_i,m^l_{i,t-1},o^l_{i,t-1})$
    \ENDFOR
\ENDFOR

\RETURN Final memory state $m^l_{i,T}$ and opinion $o^l_{i,T}$ of each agent
\end{algorithmic}
\end{algorithm}

However, owing to the potentially vast volume of interactions, storing all of them in detail is impractical.
To address this challenge, we implement a dual memory system for each agent, comprising a long-term memory $m^l_i$ and a short-term memory $m^s_i$. 
The long-term memory compresses and stores a summarized history of past interactions, while the short-term memory reflects and summarizes conversations from the current day. 
At the end of each day, agents reflect on these interactions, allowing their opinions to evolve.
The function prompt for short-term memory $f^s_m$ is as:
\vspace{-2mm}
\begin{tcolorbox}[colback=gray!20, left=1mm, right=1mm, top=1mm, bottom=1mm] 
Summarize the opinions you have heard in a few sentences, including whether or not they believe in the news.
\end{tcolorbox}
\vspace{-2mm}

\noindent and long-term memory prompt $f^l_m$ is as:
\vspace{-2mm}
\begin{tcolorbox}[colback=gray!20, left=1mm, right=1mm, top=1mm, bottom=1mm] 
Recap of previous long-term memory, today's short-term summary, please update long-term memory by integrating today's summary with the existing long-term memory, ensuring to maintain continuity and add any new insights.
\end{tcolorbox}
\vspace{-2mm}

The short-term memory is cleared at the end of each day to accommodate new interactions.
This approach achieves a balance between retaining crucial historical context and managing the volume of daily interaction data.

\textbf{Reasoning for Opinion.}
A significant difference from previous approaches involves using text descriptions to simulate each individual's perspective on fake news, providing a richer and more nuanced explanation. 
Intuitively, the evolution of these opinions is shaped by multiple factors such as personal traits, education level, social interactions, and individual reasoning processes. 
We adopt the tweet format for agents to express their opinions, as our preliminary experiments have shown that this format encourages succinct and precise statements. 
The updated prompt $f_o$ is generally as follows:
\begin{tcolorbox}[colback=gray!20, left=1mm, right=1mm, top=1mm, bottom=1mm] 
You are simulating a real person with [trait] and [education level]. 
Given your [previous personal opinion] and the new information in your [long memory], update your opinion. 
Compose a tweet expressing your opinion. 
Use 0 for disbelief and 1 for belief to indicate your opinion. 
Provide reasoning behind your tweet and explain the rationale for your belief.
\end{tcolorbox}
The overall dynamic opinion agent algorithm is shown in Algorithm~\ref{alg:generative_agent}.

\begin{table*}[tb]
\centering
\small
\begin{tabularx}{0.9\textwidth}{>{\centering\arraybackslash}p{3.8cm}|>{\centering\arraybackslash}p{1.8cm}|>{\centering\arraybackslash}p{2cm}|>{\centering\arraybackslash}p{1.8cm}|>{\centering\arraybackslash}p{1.8cm}|>{\centering\arraybackslash}p{1cm}|>{\centering\arraybackslash}p{1cm}}
\toprule
\textbf{Settings} & \textbf{Belief Average$\downarrow$} & \textbf{Belief Variance} & \textbf{Infection Rate$\downarrow$} & \textbf{Recovery Rate$\uparrow$} & \textbf{Peak Rate$\downarrow$} & \textbf{Half Rate$\uparrow$} \\
\hline
Politics Topic & 1.000 & 0.000 & 2.000 & 0.000 & 0.167 & 0.033 \\
Science Topic & 0.433 & 0.246 & 0.867 & 0.333 & 0.500 & \textgreater 1 \\
\hline
Skeptical Trait & 0.467 & 0.249 & 0.933 & 0.400 & 0.433 & \textgreater1 \\
Credulous Trait & 0.867 & 0.116 & 1.733 & 0.133 & 0.433 & 0.167 \\
\hline

No Official & 1.000 & 0.000 & 2.000 & 0.000 & 0.200 & 0.067 \\
 Single Official Declaration & 0.933 & 0.062 &1.867 &0.067 & 0.500 & 0.100 \\
 Multiple Official Declarations & 0.900 & 0.090 & 1.800 & 0.200 & 0.400 & 0.200 \\
\bottomrule
\end{tabularx}
\caption{Comparative analysis of fake news evolution across various settings, including differences in topics, traits, and intervention strategies.
Upward or downward arrows represent better control of fake news.}
\label{main}
\end{table*}

\subsection{Agent Interaction Simulator}
Alongside the DOA module, agents form a social network, allowing us to calculate the number of individuals in different groups at the macro level.
We adopt a modified SIR (Susceptible-Infectious-Recovered) model, where agents can transition between being \textit{susceptible} to fake news, becoming \textit{infected} by propagating it, and then being considered \textit{recovered} after the misinformation is corrected. 
However, unlike the traditional SIR model~\cite{zhu2017rumor}, our recovered agents can become infected again due to the dynamic nature of people's opinions. 
This aspect of our model differs from both the SIS (Susceptible-Infectious-Susceptible) model~\cite{kimura2009efficient} and the standard SIR model, where recovered individuals do not get infected again.

In the traditional SIS model~\cite{kimura2009efficient}, the simulation formulas for the infected and susceptible population are presented. 
If we modify our model by changing the `recovered' label to `susceptible', our model becomes equivalent to the SIS model, allowing us to apply the same formulas in our simulation.
The key formulas used in the SIS model are differential equations that describe how the numbers of susceptible and infectious individuals change over time:
\begin{equation}
\frac{dS}{dt} = -\beta \cdot S \cdot I + \gamma \cdot I,
\end{equation}
\begin{equation}
\frac{dI}{dt} = \beta \cdot S \cdot I - \gamma \cdot I,
\end{equation}
where $\frac{dS}{dt}$ and $\frac{dI}{dt}$ are the rates of change of susceptible and infectious individuals over time, respectively.
$\beta$ is the transmission rate, representing the probability of transmission per contact between a susceptible and an infectious individual.
$\gamma$ is the recovery rate, representing the rate at which infectious individuals recover and become susceptible again.
If $\beta$ is high and $\gamma$ is low, the disease spreads rapidly and infects a large portion of the population. 
Conversely, if $\beta$ is low and $\gamma$ is high, the disease spreads slowly and may die out.

\begin{figure*}[h]
    \centering
    \includegraphics[width=1\linewidth]{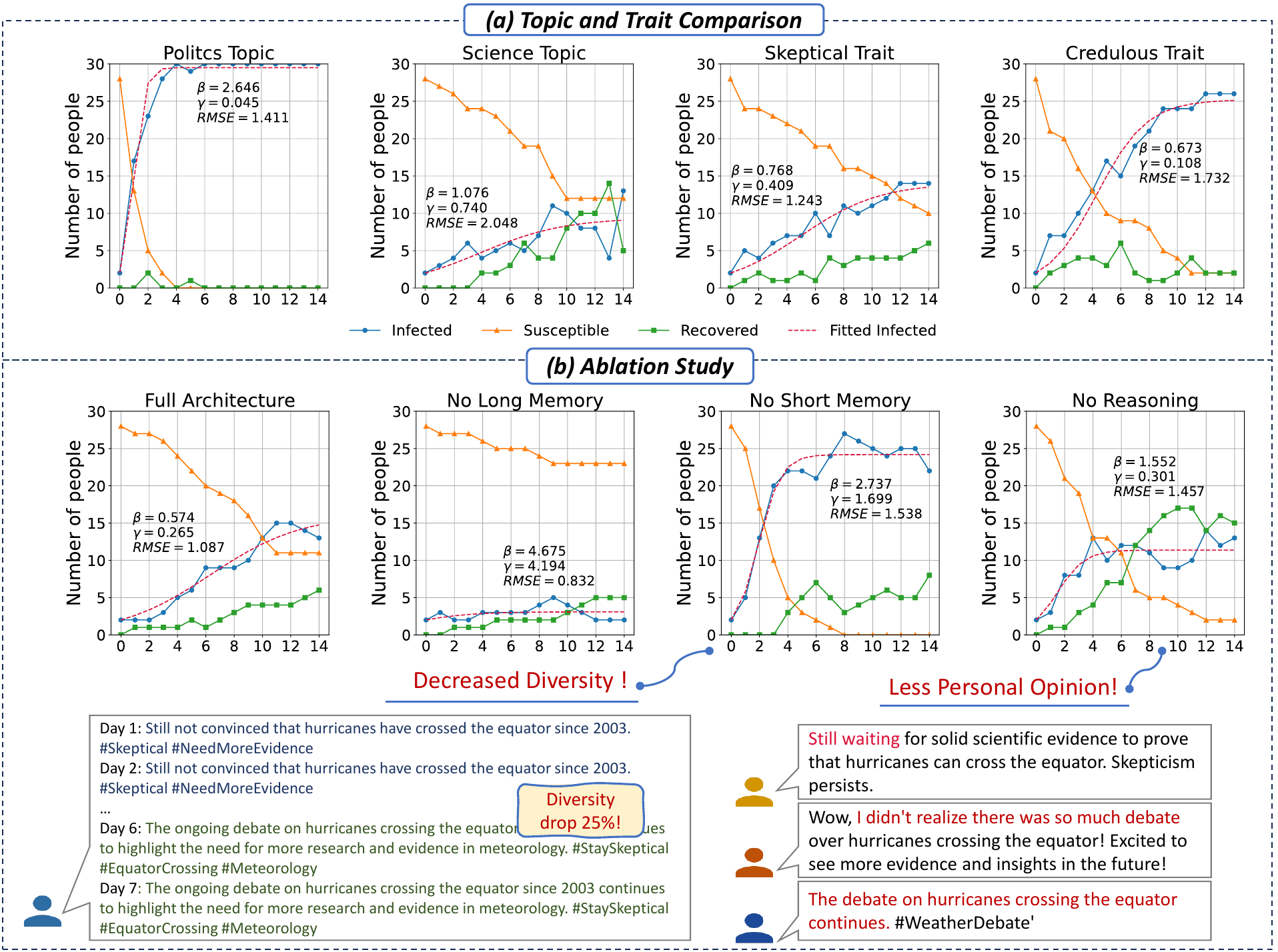}
    \caption{Dynamic group population number changes in terms of different topics and traits, with an accompanying fitting curve based on the SIS model.
    The red dashed line represents the results of the SIS model fitting, where $\beta$ is the transmission rate and $\gamma$ is the recovery rate.
    }
    \label{fig:curve}
\end{figure*}

\textbf{Intervention.}
In the fake news evolution, a critical feature is the intervention mechanism, activated when an authoritative entity propagates clarifications about fake news.
To simulate such events, our AIS also introduces a new agent designed to represent an official spokesperson.
The official agent will issue official refutations to all other agents on designated days to combat the propagation of fake news:
\begin{tcolorbox}[colback=gray!20, left=1mm, right=1mm, top=1mm, bottom=1mm] 
As the official spokesperson, I hereby issue a formal statement of refutation regarding the recent news report circulated on various media platforms concerning [topic].
\end{tcolorbox}

The full version of the prompt is in Appendix A.
We closely monitor the interactions and influence of this spokesperson to evaluate their impact on the agents' beliefs and measure how effectively they can stem the tide of the fake narrative in our simulated environment.

\section{Experiments}

\subsection{Implementation Details}

We use a Python script to operationalize our simulation FPS. 
The LLM used is gpt-3.5-turbo-1106 accessed via OpenAI API calls. 
Agents and the world they live in were defined using a Python library Mesa~\cite{kazil2020utilizing}.
Their names are selected using the names-dataset library, and ages are randomly selected from 18 to 64.
Agent traits are based on the Big Five traits typically used in psychology~\cite{barrick1991big}, with each agent having a 50\% chance of possessing a positive or negative version of each trait.
We prove that our framework can be applied on different backbones and provide API cost in Appendix B and C. 

\subsection{Metrics}
At the macro level, we can track the number of people in the Infected, Susceptible, and Recovered groups and generate trajectories that visually represent the propagation and containment of fake news within the network.
Furthermore, the fitted parameters, such as the recovery rate $\gamma$ and the transmission rate $\beta$, can also be used to demonstrate the dynamics of the network.

Finally, we design additional statistical metrics to give an intuitive understanding of the propagation of fake news. 
The `Belief Average' measures the group's mean belief in fake news at the simulation's end, while the `Belief Variance' assesses belief diversity. We track the `Infection Rate', based on the final infected count over the simulation duration, and the `Recovery Rate', calculated from the recovered count. 
`Peak Rate' reflects the maximum infection level during the simulation, and `Half Rate' is the time required for half of the group to become infected.

\subsection{Macro-level Observation}

\textbf{Topic Comparison.}
We conduct experiments on six different topics such as politics, science, and terrorism.
Generally, we found that fake political news propagates faster than false news about terrorism, science, or financial information, which is consistent with previous research~\cite{vosoughi2018spread}. 
Here, we selected two topics for which the group curve and fitting results are shown in Figure~\ref{fig:curve}.
Other figures and details can be found in Appendix D.
It is evident that the number of infected individuals for political news grows fast, reaching its peak in just four days. 
The group quickly forms a firm opinion, uniformly believing the fake news without change. 
In contrast, for the science topic, infected number grows more slowly, fluctuating around 10 people. 
Meanwhile, the number of recovered individuals increases, indicating that people tend to form a stable opinion that is skeptical of the fake news. 
The growth comparison can also be demonstrated by the fitted $ \beta$  and $\gamma$ parameters, where the $ \beta$  for the politics topic is twice as large, and the $\gamma$ is one-tenth.
This demonstrates that it is easier for agents to identify false news in science compared to politics.
The good alignment between our simulated number and the classic SIS model also verifies the accuracy and reliability of our simulation approach.
Table~\ref{main} presents more statistical results. 
In politics, the belief average is higher with less variance, accompanied by a larger infection rate and a zero recovery rate.
The topics of terrorism and financial information exhibit similar trends to science as shown in Appendix.

\textbf{Trait Comparison.}
The Big Five personality traits framework categorizes human personality into five broad dimensions: Openness, Conscientiousness, Extraversion, Agreeableness, and Neuroticism. 
Studies~\cite{ibrahim2022effects,mirzabeigi2023role} have found that individuals with high agreeableness and high neuroticism are more likely to believe rumors than those with low agreeableness and low neuroticism.
Therefore, we conduct a comparative study in which the three other traits are still randomly sampled, but for the two selected traits, one group's traits are only sampled from the positive choices, meaning high agreeableness and high neuroticism, which we denote as the `Credulous Trait'. 
Conversely, we call the opposite group, characterized by low agreeableness and low neuroticism, the `Skeptical Trait'.

From Figure~\ref{fig:curve}, we can see that the infection curve for the credulous trait rises more rapidly compared with the skeptical trait. 
The number of recovered individuals shows a slight upward trend, which aligns with the intuition that skeptical individuals tend to question external viewpoints and maintain more consistent opinions. 
In contrast, the recovery number for the credulous trait is always around 2, indicating that they are more prone to changing their opinions frequently.
The statistic numbers in Table~\ref{main} also confirm these observations.

All the above results align with previous findings, demonstrating the effectiveness of our framework.

\subsection{Micro-level Observation}

\begin{figure*}[htbp]
    \centering
    \includegraphics[width=1\linewidth]{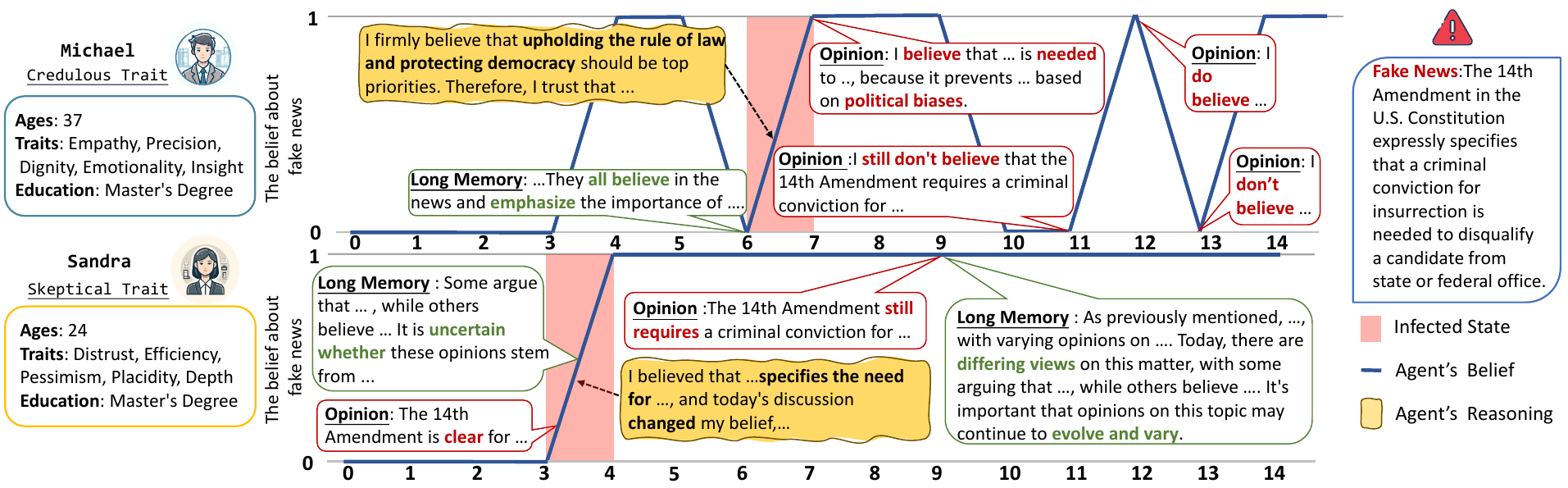}
    \caption{Micro-level case study of two people of different traits.
    Michael, possessing a credulous nature, frequently changes his opinions, whereas Sandra, being skeptical, tends to maintain a consistent view.
    }
    \label{fig:case}
    \vspace{-4mm}
\end{figure*}

Figure~\ref{fig:case} provides a comparative micro-analysis of two individuals' responses to fake news.
Michael, characterized by his credulous nature, as demonstrated by `empathy' (high agreeableness) and `emotionality' (high neuroticism), is prone to being influenced by others. 
For instance, on the sixth day, he altered his perspective to align with others, because others `\textit{all believe}' in his memory records. 
Furthermore, he frequently changes his opinions, with his stance on fake news shifting back and forth over several periods of 3 to 14 days. 
Notably, his reasons for these opinion changes can be traced back to a well-reasoned thought process. 
For example, on the sixth day, his decision to believe in the news was influenced by his conviction in principles like `\textit{I firmly believe in upholding the rule of law and protecting democracy}'.
This indicates that the opinion shifts in our FPS are grounded in thoughtful reasoning.
In contrast, Sandra possesses a skepticism trait characterized by `distrust' (low agreeableness) and `placidity' (low neuroticism), which makes her more skeptical and composed.
Throughout the 15 days, She seldom changes her opinion.
On the ninth day, despite encountering different viewpoints and storing phrases like `\textit{differing views on this matter}' and `\textit{opinions on this topic may continue to evolve and vary}' in her memory, Sandra remained unchanged in her opinion.

\begin{figure*}[htbp]
    \centering
    \resizebox{!}{0.19\textheight}{
        \includegraphics{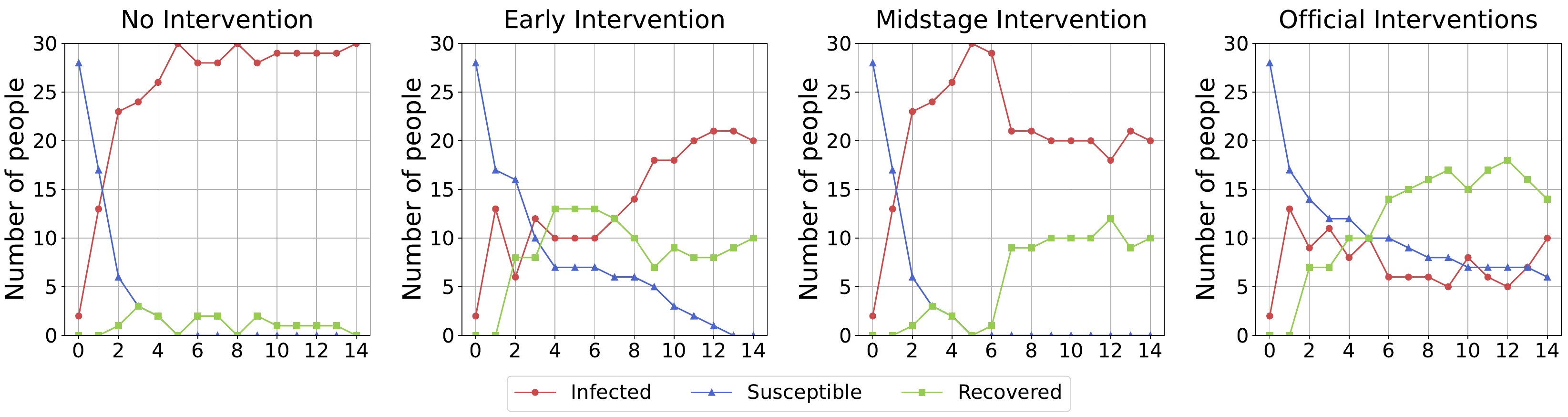}
    }
    \caption{Comparison of models with different office agent intervention strategies. It can be seen that early and reasonably frequent regulation on fake news can lead to a significant reduction in its propagation and influence.}
    \label{fig:official}
    \vspace{-4.5mm}
\end{figure*}

\section{Analysis and Discussion}

\subsection{Fake News Intervention}

Remember that in the AIS section, we integrate an official agent to fight against fake news.
Here, we first study the proper intervention strategy of the official agent.
We study political topics, as they are most impacted by fake news, posing significant challenges for fact-checking and debunking.

\textbf{Intervention Strategy.}
Firstly, we introduce the official agent into our model on both the first day and the seventh day.
On the first day, the fake news has not yet propagate widely, but by the third day, many already believe it.
As illustrated in Figure~\ref{fig:official}, each timing choice offers distinct advantages.
Introducing the official agent at the beginning can substantially reduce the number of people initially believing in fake news.
However, as time progresses, people may revert to believing in the fake news due to a forgetting mechanism.
On the other hand, introducing the official agent on the seventh day demonstrates a more sustained effect, with a gradual and consistent decline in the number of affected individuals. 
However, since the number of people influenced by fake news is already high by the seventh day, a substantial portion of the population remains affected despite the intervention.

This leads to our second experiment, where we investigate the frequency of introducing official agent necessary to maintain fake news propagation at a manageable level. 
We tested scenarios of releasing official news daily and every three days as shown in `Official Intervention'. 
Results show that there is no significant difference between these two approaches, indicating that fact-checking news can be released at intervals without compromising its effectiveness in controlling fake news at a reasonable cost. 

\textbf{Chronic Believers in Fake News.}
Despite the optimistic outlook on interventions against fake news, we observe that some individuals still become infected. 
Our analysis of the data reveals that approximately 50\% of the infected group remains infected throughout, indicating that even official interventions have a limited impact on changing their beliefs.
Upon examining these cases, we find that the infected individuals vary in age and education level. 
However, they share common characteristics such as `high agreeableness'. These traits align with findings from our previous trait study.
This observation suggests that interventions against fake news might be more effective if they are tailored to address these specific traits.

\subsection{Ablation Study}

We chose a science topic to demonstrate the effectiveness of our model's components, including long-term memory, short-term memory, and reasoning, as shown in Figure~\ref{main}(b). This setting illustrates a balanced dynamic of opinion change.

Firstly, when long-term memory is removed, the influence of short-term interactions from today alone is insufficient to persuade people to change their opinions. 
In this scenario, the simulation fails to produce effective interactions.
The absence of short-term memory hinders the reflection and consolidation of new information, leading to the formation of long-term memory that merely accumulates daily opinions without deep analysis. 
This process results in monotonous and undiversified opinions.
To quantify this, we employed the diversity metric~\cite{li2015diversity}. 
Our analysis reveals that the average opinion diversity in the FPS without short-term memory is only three-quarters of the diversity score compared to the full model.
The full score can be found in Appendix E.
Finally, we examined the impact of removing the reasoning process in the opinion update mechanism. 
This alteration has the least effect on overall performance, as it retains the main components of our framework. 
However, a closer examination of the generated opinions reveals that the agents' opinions resemble mere descriptions of the opinions they receive, lacking the advanced function of thoughtful reasoning or expressing their own opinions.

\section{Conclusion}
In this study, we present the first LLM-based simulation framework for fake news research, incorporating short-term and long-term memory, along with reasoning processes that mimic human cognition. 
Our simulations offer not only micro-level observations that align with previous studies on fake news topics and traits but also correspond with earlier numerical studies on macro-level simulations.
It also includes an intervention mechanism to curb the propagation of fake news, providing practical strategies for real-world monitoring. 
This innovative approach aims to advance research in the field of fake news analysis and mitigation.

\bibliographystyle{named}
\bibliography{ijcai24}

\end{document}